\documentclass[amsmath,amssymb,aps,floatfix,prd,a4paper]{revtex4}
\usepackage{epsfig}%
\usepackage{amsmath}%
\usepackage{graphicx}
\usepackage{amsmath}
\usepackage{amsfonts}
\usepackage{amssymb}
\usepackage{physics}
\usepackage{changepage}
\usepackage{caption}
\usepackage{verbatim}

\newcommand{\beq}{\begin{equation}}
\newcommand{\eeq}{\end{equation}}

\newcommand{\ave}[1]{\left\langle #1 \right\rangle}

\begin{document}

\title{The energy dependence of  the multiplicity moments at the LHC}
\author{G. R.  Germano}
\email[e-mail: ]{ guilherme.germano@usp.br}
\author{F. S. Navarra} 
\email[e-mail: ]{ navarra@if.usp.br}
\affiliation{Instituto de F\'{\i}sica, Universidade de S\~{a}o Paulo, 
  Rua do Mat\~ao, 1371, CEP 05508-090, Cidade Universit\'aria, 
  S\~{a}o Paulo, SP, Brazil.}

\begin{abstract}

In this work, from the experimental data we evaluate the first C-moments
of the multiplicity distributions recently measured in proton-proton
collisions at the 
LHC and compare them  with the predictions of two models: the
Kharzeev-Levin model and the Bialas-Praszalowicz model. We divide the
data into three sets according to their phase space coverage: I:
$p_T > 100 $ MeV and $|\eta|< 0.5$; II: $p_T > 100 $ MeV and $|\eta|< 2.4$
and II: $p_T > 500 $ MeV and $|\eta|< 2.4$. The mean multiplicity grows with
the energy according to a power law and the power is different for each set.
The $C_n$ moments grow continuously with the energy, slowly in set I and
faster in the other sets. Except for KL in set II, both models reproduce the
main features of the data. The negative binomial parameter $k$ decreases
continuously with the energy and there is no sign of change in this behavior.

\end{abstract}
\maketitle

\section{Introduction}

Particle production in hadronic collisions at very high energies is a very
interesting phenomenon. On the experimental side particle multiplicities
are easy to measure. On the theoretical side it is still a very challenging
subject \cite{gor}.  Most of the  particles are produced with low and medium
transverse
momentum, where perturbative QCD cannot be applied and one has to use 
phenomenological models and/or Monte Carlo event generators. In
\cite{cms11} it was shown that these generators are able to reproduce the
main features of the multiplicity distribuions but they are not yet able to
describe the data with precision, especially in the large multiplicity
region.  

Over the past twenty years the QCD based theory of particle production has
experienced a significant progress, especially because of the development
of the Color Glass Condensate (CGC) formalism \cite{cgc}.   
One of the most interesting 
predictions of the CGC is that at very high energies the multiplicity
distributions will become narrower \cite{Gelis09}. This may be observed
looking at the
behavior of the multiplicity moments $C_n$, which should decrease with the
energy. Ten years ago this prediction was confronted with the LHC data in
\cite{praza11}, where a careful study of the moments was carried out. The
conclusion was that the moments were continuously growing with the energy
with no sign of change in this trend. Since then, new data appeared, taken
at much higher energies, and it is time to check if there is anything
special happening to the moments.  Here it is important to mention that
in the most recent experimental papers the multiplicity distributions  
$P(n)$ were presented but the moments $C_n$ were not.  The first  goal   
of this work is thus to compute the moments from the multiplicity tables,
which we took from the hepdata.net databasis (the corresponding links are
given right after the related articles in the reference list). Then we will
compare them with the predictions of two simple models. The first one, which
we call Kharzeev-Levin (KL) was 
proposed in \cite{lubi} and used in \cite{kale} to compute the moments of 
charged particle multiplicity distributions. This model is based on the 
Balitsky-Kovchegov (BK) equation with fixed dipole sizes. In the second
model, called here Bialas - Praszalowicz (BP) Model \cite{praza11,bialas10},   
multiparticle production is described by a probability distribution,
which is a superposition of a distribution of the number of sources 
and a Poisson distribution describing particle emission from each source. 
In the KL model there is always only one source, whereas in the BP model the
number of sources grows with energy and can be large. In this aspect, these
models are complementary.

The predictions of KL and BP models will be compared with the most recent  
data from the LHC on non single diffractive pp collisions, which can be
grouped into three sets:
\begin{itemize}
        \item Set  I: $p_{T} > 100 $ MeV, $|\eta|<0.5$, and energies 
        $\sqrt{s} =$ 900, 2360 and 7000 GeV, from CMS \cite{cms11}.
        
        \item Set II: $p_{T} > 100 $ MeV, $|\eta|<2.4$, and energies
          $ \sqrt{s} = $  900,  7000 and 8000 GeV from ALICE \cite{alice17};
          $ \sqrt{s} = $  8000 GeV
        \cite{atlas16a}  and  13000 GeV \cite{atlas16b} from ATLAS.
        
        \item Set  III: $p_{T} > 500 $ MeV, $|\eta|<2.4$,  and energies
        $\sqrt{s} = $ 900, 7000 GeV  \cite{atlas11},
        8000 GeV ($|\eta|<2.5$) \cite{atlas16a}, 
        13000 GeV ($|\eta|<2.5$) \cite{atlas16c} from ATLAS  and
        13000 GeV ($|\eta|<2.4$) \cite{cms18} from CMS.
        In \cite{atlas11} there are also data from the run at  
        $\sqrt{s} = $ 2.36 TeV, but these data were truncated at     
        large $n$ due to a problem with the detector, as explained in 
        the paper. Large $n$ plays an important role in the
        evaluation of C-moments, making these data unreliable for our
        analysis.
\end{itemize}

These data sets may contain particles produced through different production
mechanisms. The particles measured in set I are produced mainly from gluons;
those measured in set II come also from the fragmentation region (larger
rapidities) and hence are produced also from the valence quarks.
Due to the larger transverse momentum cut-off, set III contains more particles
which are produced perturbatively. These differences might lead to a different
behavior of some observables. In previous analyses \cite{cms11}, it has
been observed that multiplicity distributions in set I satisfy the
Koba-Nielsen-Olsen (KNO) scaling, whereas those in set II do not. 

\section{The Models} 

\subsection{The Kharzeev - Levin Model }

In Ref. \cite{lubi,kale}, the authors developed a model  for     
multiplicity distributions based on the BK equation, which
we will call KL model.  They propose the following
evolution equation for the parton multiplicity distribution $P_n$:
\beq
\frac{d P_n(Y)}{d Y} \, = \, - \, \Delta \, n \, P_n  \, + \,
(n-1) \, \Delta \, P_{n-1} (Y)
\label{kl}
\eeq
which has the simple solution:
\beq
P_n(Y) \,  = \, P_{KL}(n) \, = \, 
e^{-\Delta Y} \left(1 - e^{- \Delta Y} \right)^{n - 1}
\label{klpn}
\eeq
where $Y = ln(1/x)$ and $\Delta$ is the BFKL Pomeron
intercept, $\Delta\, = \,  4 \,\,ln \,\,  2  \,\,  \bar{\alpha}_s$ with 
$\bar{\alpha}_s = \alpha_s N_c / \pi$. From the above expression we obtain
the mean multiplicity: 
\beq
\langle n  \rangle  = \sum_n n \, P(n) = e^{\Delta Y} =
\left( \frac{1}{x} \right)^{\Delta}
\label{nbfkl}
\eeq
The variable $x$ is defined here as in \cite{lelu,babi}:
\beq
x = \frac{q_0^2}{s}
\label{defex}
\eeq
where $q_0$ is a constant. 
Inserting (\ref{defex}) into (\ref{nbfkl}) we obtain:
\beq
\langle n  \rangle  =  \left( \frac{s}{q_0^2} \right)^{\Delta} 
\label{nf}
\eeq
The energy scale $q_0$ can be a mass or the average transverse momentum and
hence it might be different for different data sets, but it should not
depend on the collision energy $\sqrt{s}$. In what follows we will fix 
$q_0$ and $\Delta$ from  the fit of the available experimental data.

\subsection{The Bialas - Praszalowicz  Model }

In the BP model the multiplicity distribution is given by: 
\begin{equation}
P_{BP}(n)=%
{\displaystyle\int\limits_{0}^{\infty}}
dt\,F(t)\,e^{-\bar{n}t}\frac{(\bar{n}t)^{n}}{n!}.
\label{convol}
\end{equation}
where $t$ is a fraction of the average multiplicity, and $F(t)$ the
distribution of sources that contribute a fraction $t$ to the multiplicity
probability $P_{BP}(n)$. The function $F$ is normalized:
\begin{equation}%
{\displaystyle\int\limits_{0}^{\infty}}
dt\,F(t)=%
{\displaystyle\int\limits_{0}^{\infty}}
dt\,t\,F(t)=1.
\label{normaliz}%
\end{equation}
The above equation implies that  $\left\langle n\right\rangle =\bar{n}$. 
In the BP model  $F$ is given by 
\begin{equation}
F(t,k)=\frac{k^{k}}{\Gamma(k)}t^{k-1}e^{-kt}\label{NBD}%
\end{equation}
and $P_{BP}$  turns out to be the negative binomial distribution (NBD), 
which is known to describe relatively well the data at lower energies
\cite{kittel}. Distribution (\ref{NBD}) depends on one parameter $k$,
which depends on the collision energy. The 
analysis of lower energy data shows that $k$ decreases with increasing energy.
When $k=1$ the
probability distribution $P_{\text{NBD}}$ becomes a  geometrical
distribution $P(n)=\left\langle n\right\rangle ^{n}/(1+\left\langle
n\right\rangle )^{n+1}$  When  $k$ is large ($1/k\rightarrow 0$), the
distribution $P_{\text{NBD}}$ tends to a Poisson distribution.

\section{Results}

\subsection{The mean multiplicity}\label{meanmulti}

\begin{figure}[!t]
\begin{tabular}{cc}
\centerline{
{\includegraphics[height=5.0cm]{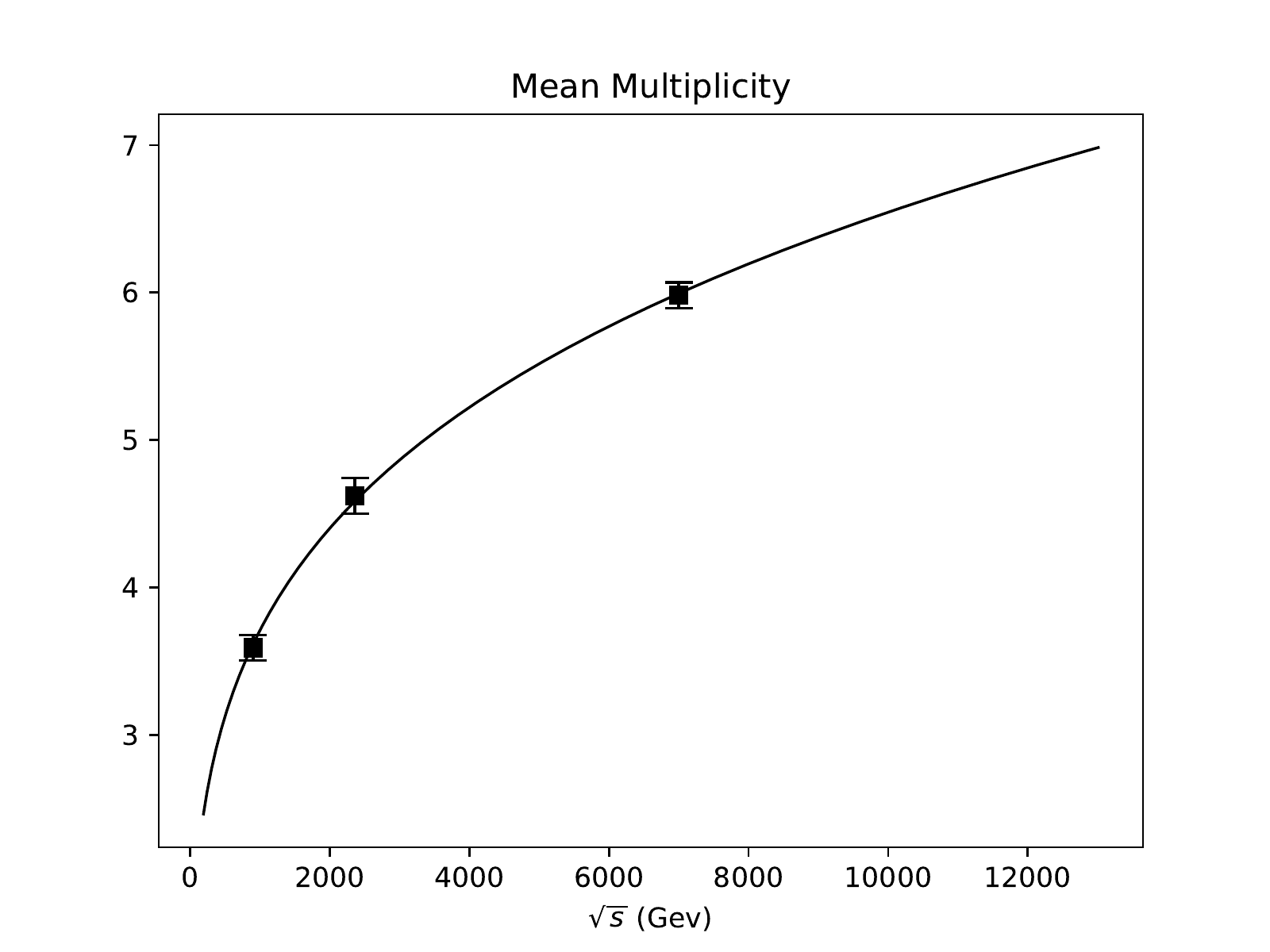}}
{\includegraphics[height=5.0cm]{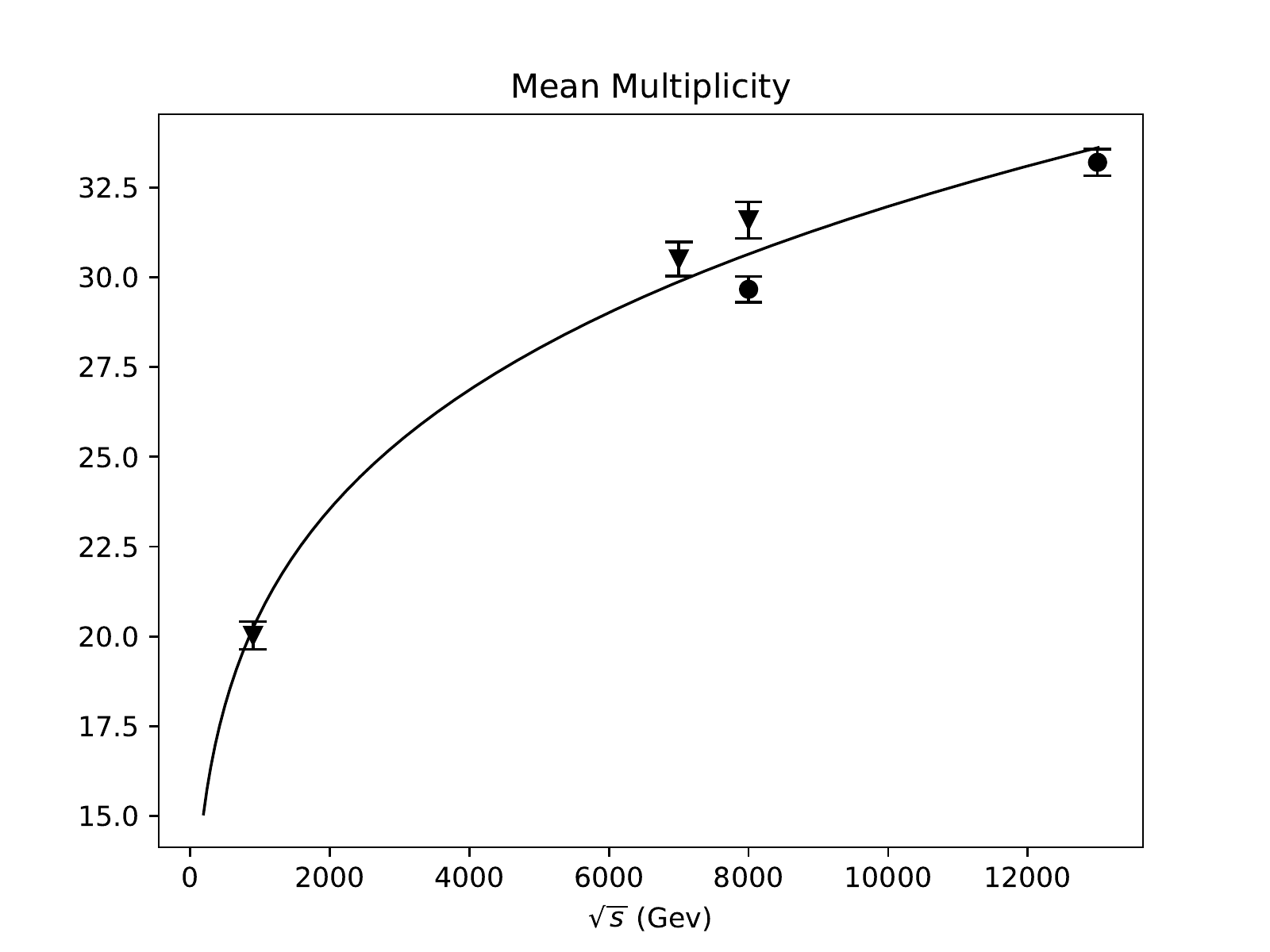}}
}
\end{tabular}\\
\begin{tabular}{cc}
\centerline{
{\includegraphics[height=5.0cm]{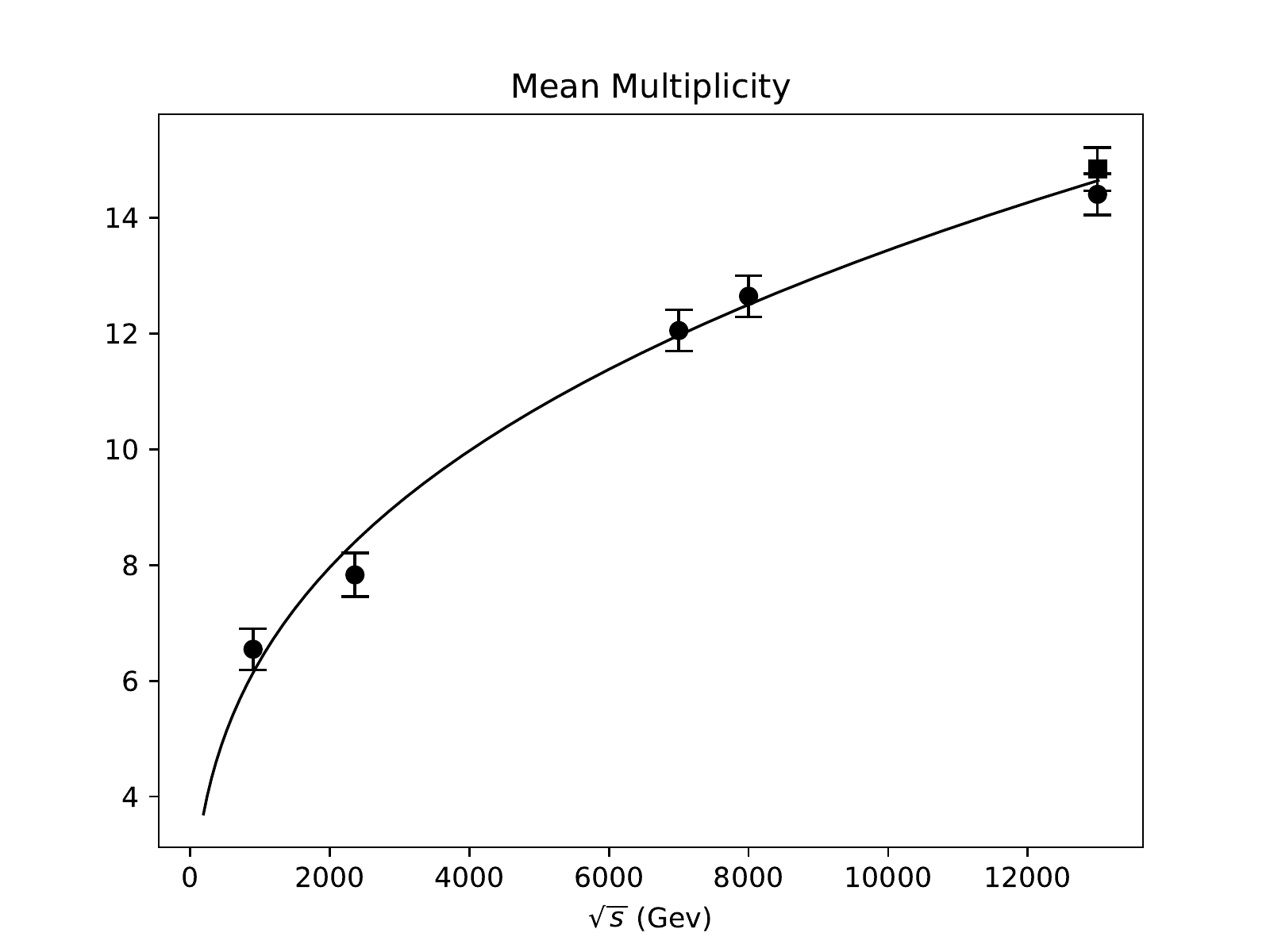}}
{\includegraphics[height=5.0cm]{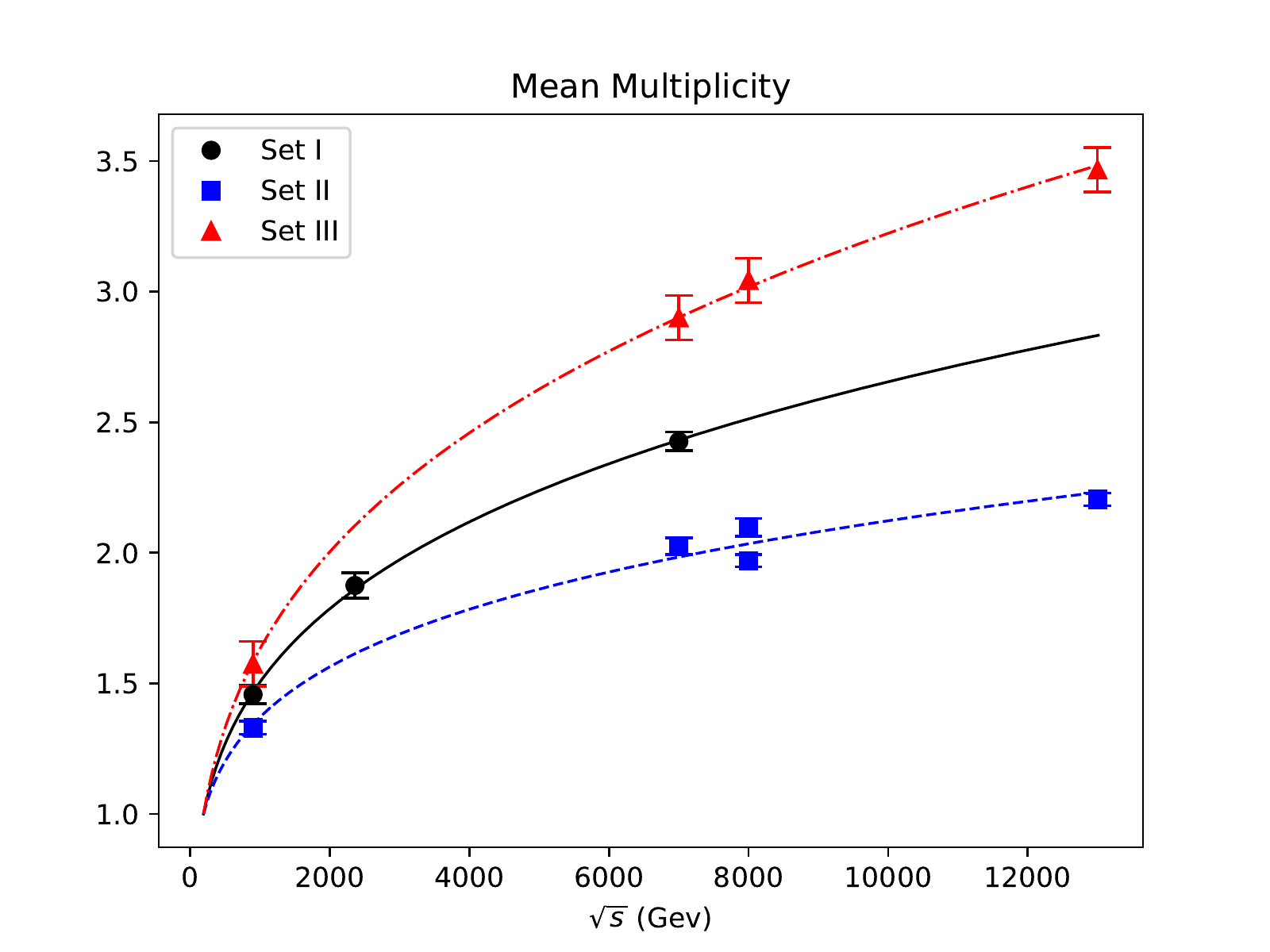}}
}
\end{tabular}
\caption{Mean multiplicities fitted with (\ref{nf}).
  a) Upper-left: Set I. Data from \cite{cms11}.
  b) Upper-right: Set II. Data from \cite{atlas16a,atlas16b} (circles) and
  from \cite{alice17} (triangles).
  c) Lower-left: Set  III. Data from
  \cite{atlas11, atlas16a, atlas16c} (circles) and \cite{cms18} (squares).
  d) Lower-right: mean multiplicities adjusted with  (\ref{nf}) for all
  sets. The points in each set were multiplied by a constant. 
}
\label{fig1}
\end{figure}

We start fitting the data on the mean multiplicity as a function of the 
energy with the help of the parametrization (\ref{nf}), as shown in       
Fig.~\ref{fig1}. The obtained values of $q_0$ and $\Delta$ are presented
in Table \ref{tab_q0delta}. As it can be seen in the figure, the data are well
reproduced by the power law (\ref{nf}). However, there is a striking difference
in the size of the parameters. For set II, the power $\Delta$ is much smaller
and the corresponding energy dependence is much weaker than in the other
sets.  
This is not an artifact of the fitting procedure and is rather a feature of the
data. To illustrate this, in Fig.~\ref{fig1}d we have have multiplied the
data and the fitting curves by appropriate constants so that the curves start
at the same point. The resulting plot shows clearly the different energy
behavior of the three sets. Since in sets II and III the rapidity coverage is
the same, the difference in the energy behavior of the mean multiplicity must
be related to the difference in the lower $p_T$ cut. As noted in the
introduction, the most natural explanation (if not the only one) for this
energy behavior is the increase of the contribution of perturbative events,
which are known to have a strong energy dependence. It is nevertheless
remarkable how a relatively modest increase in $p_T$ (which is here still far
from the typical few GeV region) can produce a visible effect.

The data of sets I and II have the same lower $p_T$ cut and very different
rapidity coverage. In set I, we observe particles produced in the central
(rapidity) region, which is dominated by gluons. In set II  there is a larger
contribution coming from the fragmentation region, where the valence quarks
play an important role in particle production. The separation of central and
fragmentation regions and its relevance for particle production was first
discussed in \cite{vp} and later implemented as a model in \cite{marb}.
In this region,  partons from 
the projectile and from the target collide in a very asymmetric kinematical
configuration. If a parton from the projetile carries a large momentum
fraction of the proton, the one from the target carries a very small fraction
of the target proton momentum. Therefore, data in this region of the phase
space are more sensitive to the low-x  QCD dynamics and to saturation effects.
As it is well known, saturation tames the growth of observables (parton
distribution functions, color dipole scattering amplitudes, hadron cross
sections, etc...) with the energy. According to these ideas, it is tempting to 
interpret the weaker energy dependence of the data of set II (as compared to
set I) as a manifestation of low x saturation effects. 

\begin{table}[h]
        \begin{center}
                \begin{tabular}{ccc}
                        \hline
                        Set & $\Delta$ & $q_{0}$ (GeV) \\
                        \hline
                        \hline
                        I & 0.13 & 6.31  \\
                        \hline
                        II & 0.05 & 0.01  \\
                        \hline
                        III & 0.16 & 4.83 \\
                        \hline
                \end{tabular}
                \caption{Parameters $q_0$ e $\Delta$.}
                \label{tab_q0delta}
        \end{center}
\end{table}

\subsection{The  $C_n$ moments}\label{cmoms}

\begin{figure}[!t]
\begin{tabular}{cc}
\centerline{
{\includegraphics[height=5.0cm]{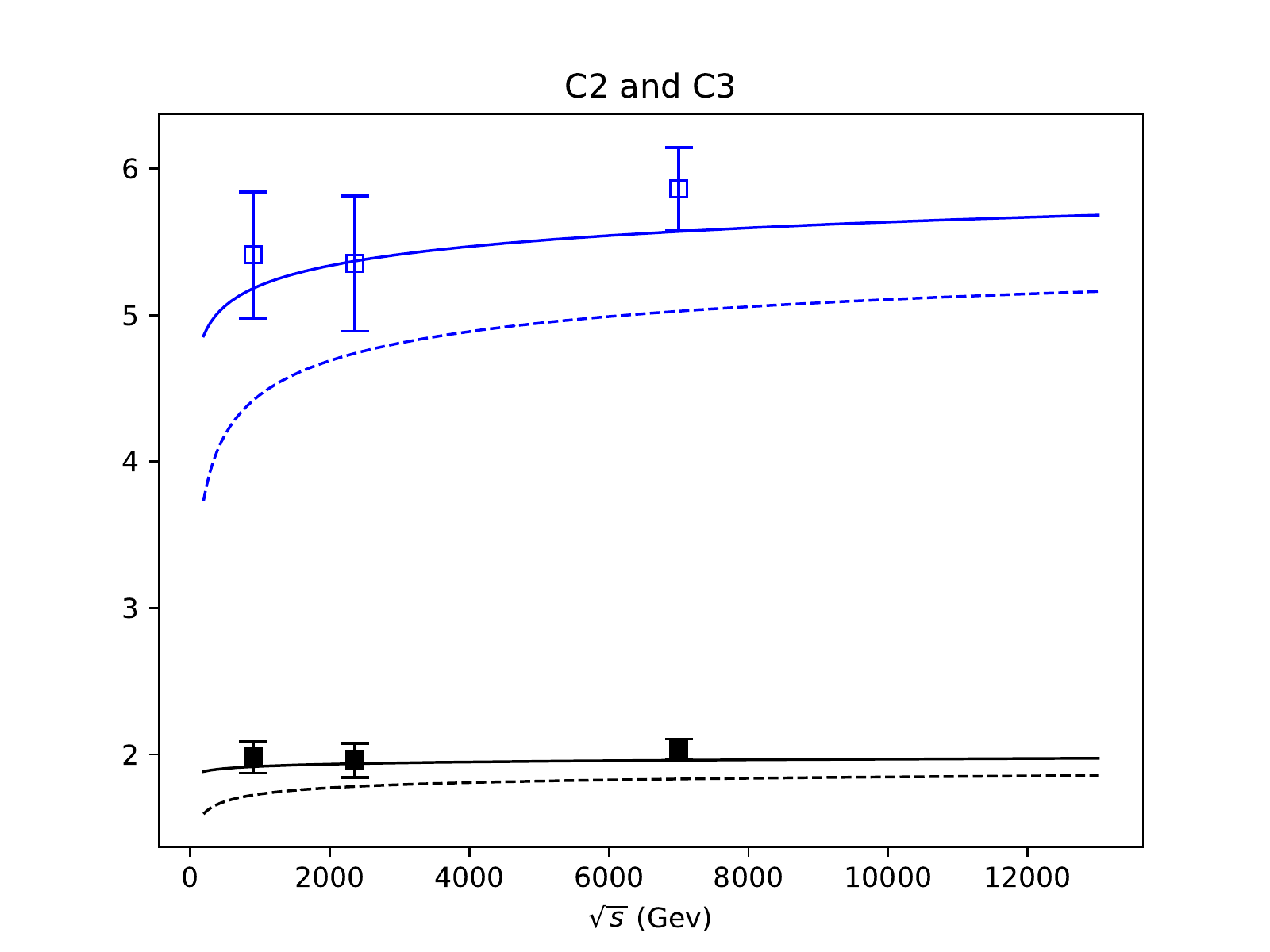}}
{\includegraphics[height=5.0cm]{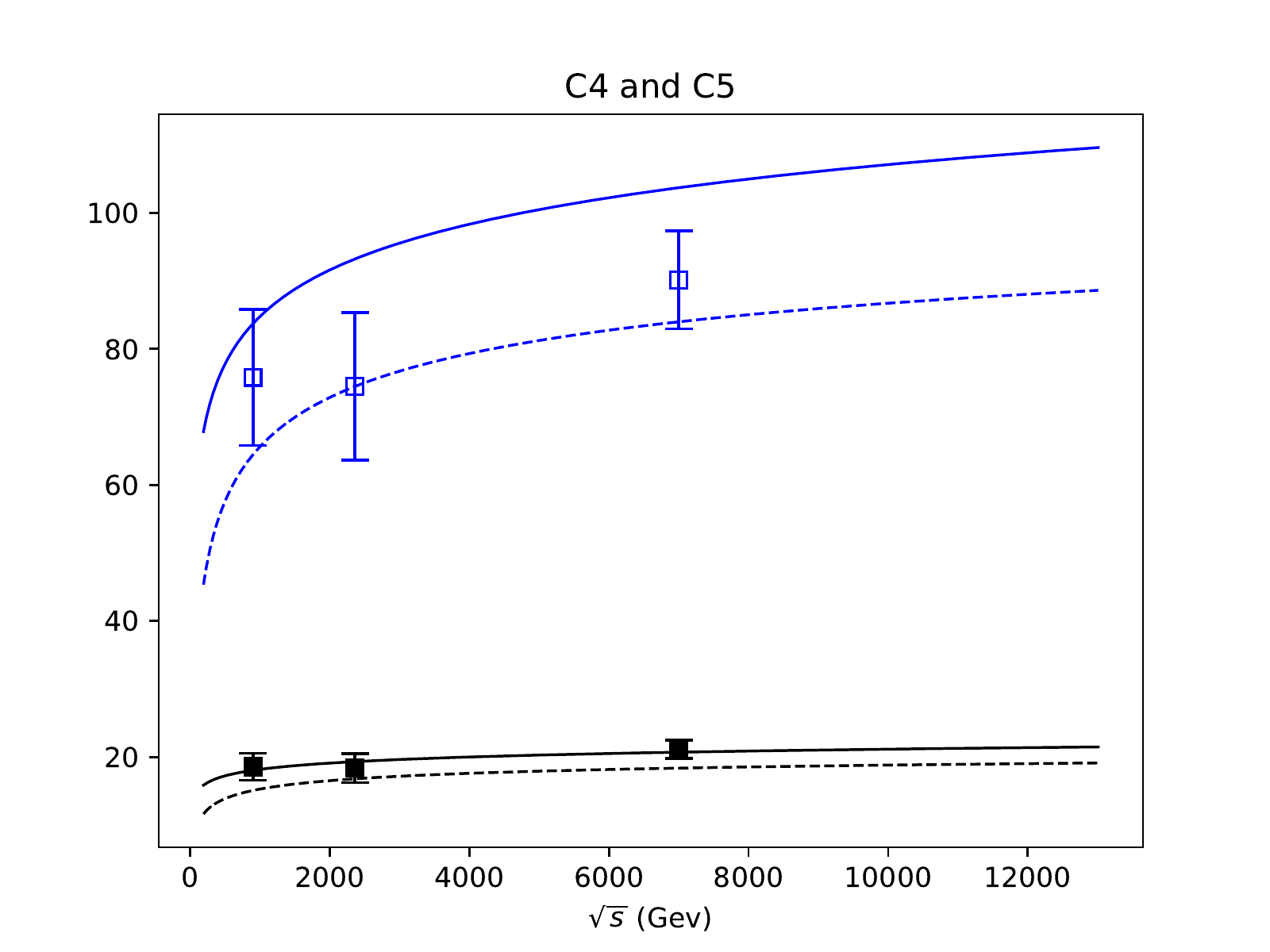}}
}
\end{tabular}  \\
\begin{tabular}{cc}
\centerline{
{\includegraphics[height=5.0cm]{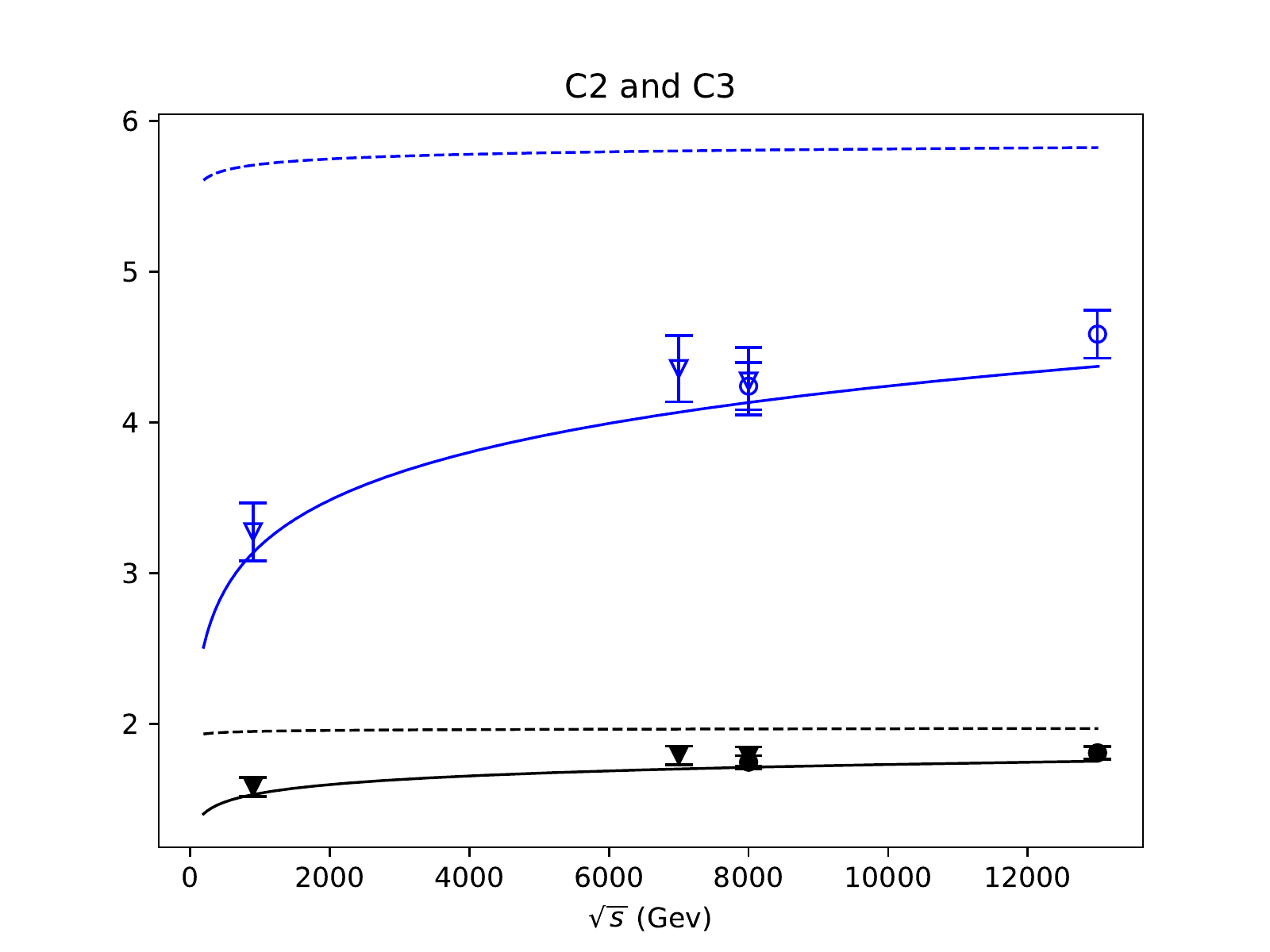}}
{\includegraphics[height=5.0cm]{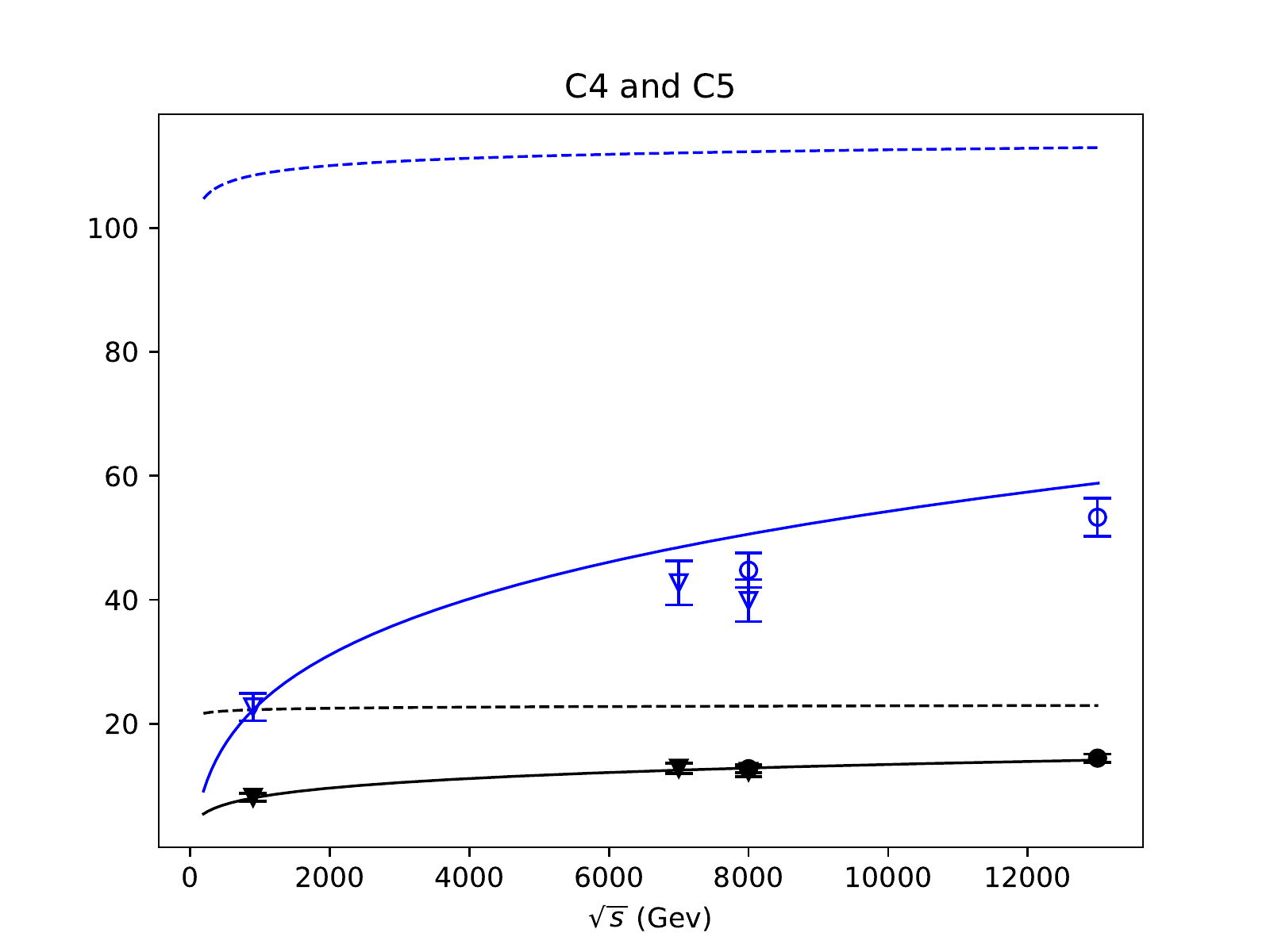}}
}
\end{tabular}\\
\begin{tabular}{cc}
\centerline{
{\includegraphics[height=5.0cm]{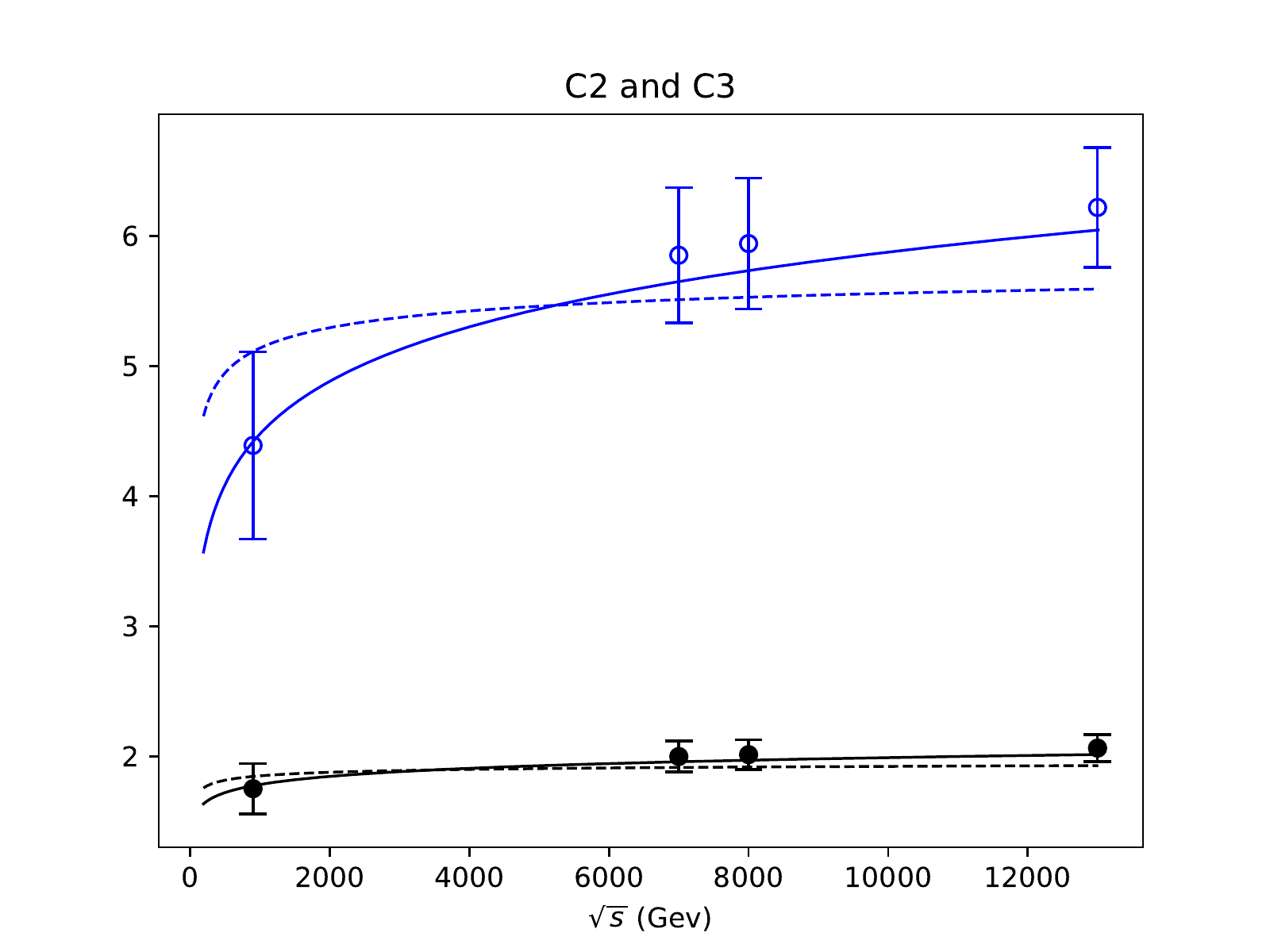}}
{\includegraphics[height=5.0cm]{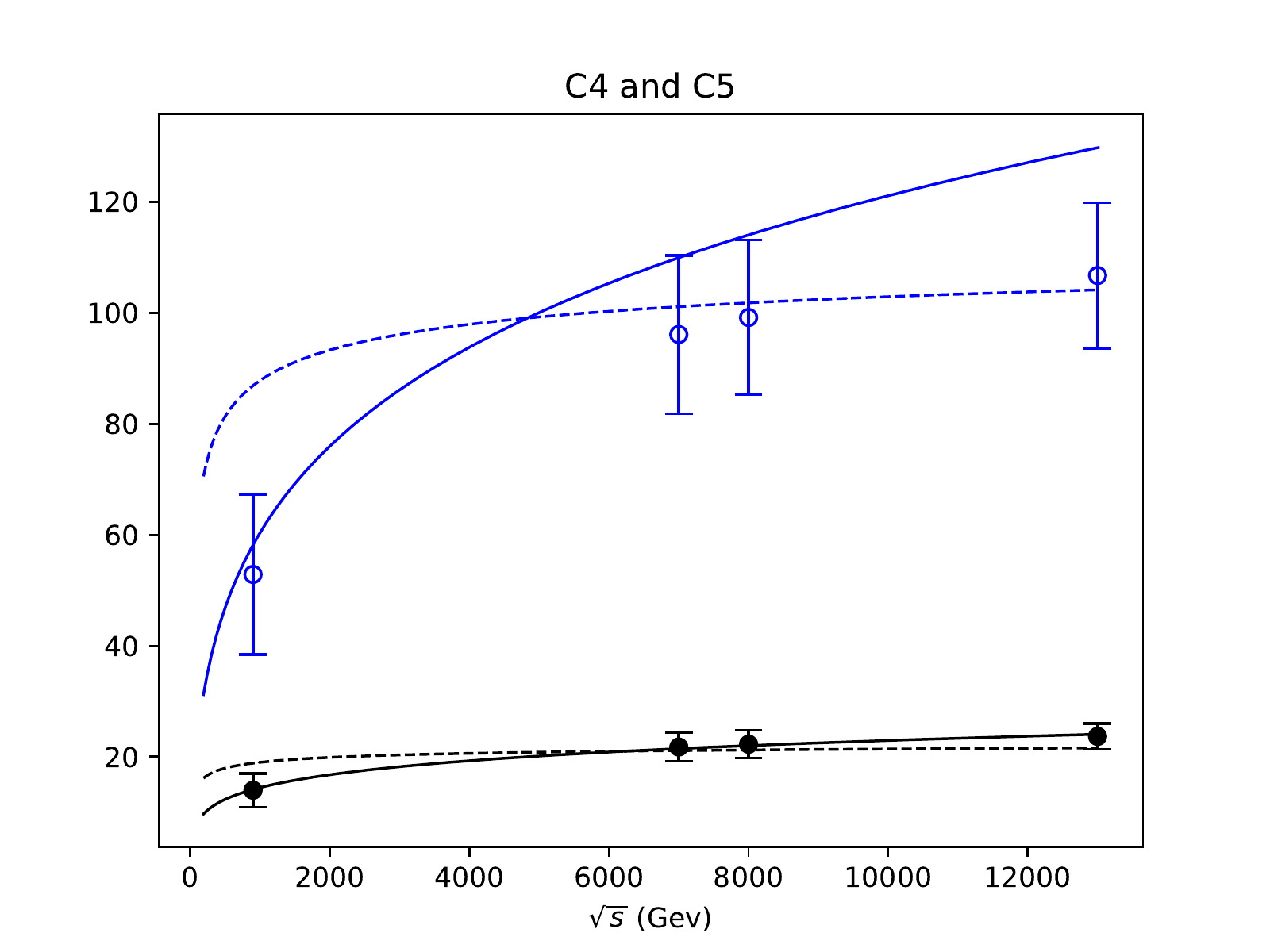}}
}
\end{tabular}
\caption{$C_n$ moments. Solid lines: BP model. Dashed lines: KL model.  
  a) Upper-left:  $C_2$ and $C_3$. Set I. Data from \cite{cms11}.
  b) Upper-right: $C_4$ and $C_5$. Set I. Data from \cite{cms11}. 
  \cite{atlas16a,atlas16b} (circles) and from \cite{alice17} (triangles).
  c) Mid-left: $C_2$ and $C_3$. Set II. Data from \cite{atlas16a,atlas16b}
  (circles)  and \cite{alice17} (triangles).
  d) Mid-right: $C_4$ and $C_5$. Set II. Data from \cite{atlas16a,atlas16b}
  (circles)  and \cite{alice17} (triangles).
  e) Lower-left:  $C_2$ and $C_3$. Set III. Data from
  \cite{atlas11, atlas16a, atlas16c}.
  f) Lower-right: $C_4$ and $C_5$. Set III. Data from
  \cite{atlas11, atlas16a, atlas16c}. 
}
\label{fig2}
\end{figure}

The moments are defined as:
\begin{equation}
\label{momdef}
C_{m}=\frac{\left\langle n^{m}\right\rangle }{\left\langle n\right\rangle^{m}}
\end{equation}
They have been measured at the LHC and explicitly given in Refs.
\cite{cms11,alice10}. For higher energies we can calculate them from the      
hepdata.net  databasis, as we did in the case of the mean multiplicities.
In the KL model we can obtain the moments from the
definition  (\ref{momdef}) and the multiplicity distribution (\ref{klpn}).
The first of  them are given  by:
\begin{align}
\label{klmoms}
C_{2} & = 2 - \frac{1}{\ave{n}}\nonumber\\
C_{3}  & =\frac{6(\ave{n}-1)\ave{n}+1}{\ave{n}^{2}},\nonumber\\
C_{4}  & =\frac{(12\ave{n}(\ave{n}-1)+1)(2\ave{n}-1)}{\ave{n}^{3}}\nonumber\\
C_{5}  & = \frac{(\ave{n}-1)(120\ave{n}^{2}(\ave{n}-1)
  +30\ave{n})+1}{\ave{n}^4}\nonumber\\
\end{align}
Since the mean multiplicities are given by (\ref{nf}), there are no free
parameters in the calculation with the KL model.  
In the BP model the moments are obtained from  (\ref{convol}) and the
first moments $C_n$ are given by: 
\begin{align}
C_{2} & =\frac{1}{\left\langle n\right\rangle }+1+\frac{1}{k}\quad\rightarrow
\quad\frac{1}{k}=C_{2}-1-\frac{1}{\left\langle n\right\rangle }.
\label{1overk} 
\end{align}
\begin{align}
\label{bpmoms}
C_{3}  & =C_{2}(2C_{2}-1)-\frac{C_{2}-1}{\left\langle n\right\rangle
},\nonumber\\
C_{4}  & =C_{2}(6C_{2}^{2}-7C_{2}+2)-2\frac{3C_{2}^{2}-4C_{2}+1}{\left\langle
n\right\rangle }+\frac{C_{2}-1}{\left\langle n\right\rangle ^{2}},\nonumber\\
C_{5}  & =C_{2}(24C_{2}^{3}-46C_{2}^{2}+29C_{2}-6)-2\frac{18C_{2}^{3}%
-34C_{2}^{2}+19C_{2}-3}{\left\langle n\right\rangle }\nonumber\\
& +\frac{14C_{2}^{2}-23C_{2}+9}{\left\langle n\right\rangle ^{2}}-
\frac{C_{2}-1}{\left\langle n\right\rangle ^{3}}.%
\end{align}
In the above expressions, we need to know $k$ to compute the moment $C_2$
and then all the other moments. Instead of choosing values for $k$, we
follow \cite{praza11} and parametrize $C_{2}$  as
\begin{equation}
  C_{2}=a+b\log(\sqrt{s}\text{[GeV]})
  \label{C2fit} 
\end{equation}
Next, we fit $C_2$ to the data, fixing $a$ and $b$, and finally we calculate
$k$ using (\ref{1overk}). 
Actually, in order to find the parameters $a$ and $b$ we fit $C_{4}$ rather
than $C_{2}$. The obtained values are listed in Table \ref{tab_aeb}.
\begin{table}[h]
        \begin{center}
                \begin{tabular}{ccc}
                  \hline
                        Set & $a$ & $b$  \\
                        \hline
                        \hline
                        I & 1.68 & 0.02  \\
                        \hline
                        II & 1.10 & 0.07  \\
                        \hline
                        III & 1.30 & 0.06  \\
                        \hline
                \end{tabular}
                \caption{Parameters $a$ and $b$ obtained from the fit
                  of data with (\ref{C2fit}), for the three sets.}
                \label{tab_aeb}
        \end{center}
\end{table}
Having determined the parameters $a$ and $b$, which are energy independent, we
can calculate all the first C moments,  compare them with data and make
predictions. This is shown in Fig. \ref{fig2}, where we compare the KL and
BP moments with the three data sets. Looking first at the data (which are put
together here for the first time) we observe that in all figures the  moments
grow with the energy. The moments from set I grow much slower and are even
compatible with a constant value. This motivated the observation made in Ref. 
\cite{cms11}, where the authors claimed that data relative to small rapidities
exhibit KNO scaling \cite{gor,kno}, while data of  larger rapidity
intervals do not. 
Comparing the moments obtained with sets II and III we see that the $C_n$'s
grow with energy in the same (strong) way.

Both models give a reasonable description of data. The KL model has less
freedom than the BP one, since in the latter, apart from $<n>$, we can also
adjust $C_2$ to the data. This flexibility yields better fits. In
\cite{praza11} the mean multiplicity was calculated from the central rapidity
density, $dN/d \eta (\eta=0)$, while here we have used (\ref{nf}). We believe
that this procedure is more accurate. The results support the BP picture,
in which  multiple sources (the number of sources follows the
distribution $F$) produce a number of particles which follows a Poisson
distribution and both distributions depend on the energy.  On the other hand
the success of the KL model would imply that particle production comes mostly
from gluon cascading resulting ultimately from the BK equation.  In view of
its simplicity, the KL model does a reasonable job with sets I and III and
fails badly with set II. In this set we have the bigger multiplicities and,
as already pointed out in \cite{kale}, with the expressions (\ref{klmoms})
we rapidly reach the asymptotic values of the ``large $<n>$'' limit. The
resulting moments are large and  very flat functions of $\sqrt{s}$,  
in contradiction with the data. It is not clear which part of the model is
wrong or incomplete. Having in mind the observations made in the case of the
mean multiplicities, we would guess that there is a missing component in
this version of the model, namely the contribution from the fragmentation
region. Increasing the lower $p_T$ cut and selecting higher $p_T$ particles,
we reduce the contribution of the fragmentation region yield (which is mostly
forward and with low transverse momentum) and improve the agreement between
the KL model and data, as is indeed seen in the comparison with set III data. 
Finally, it is worth mentioning that the discrepancy between the KL model and
the set II data is visible not only in the $C_n$ moments but also in the
entire distribution $P(n)$, as shown in \cite{nos}.

\begin{figure}[!t]
\centerline{
{\includegraphics[height=5.0cm]{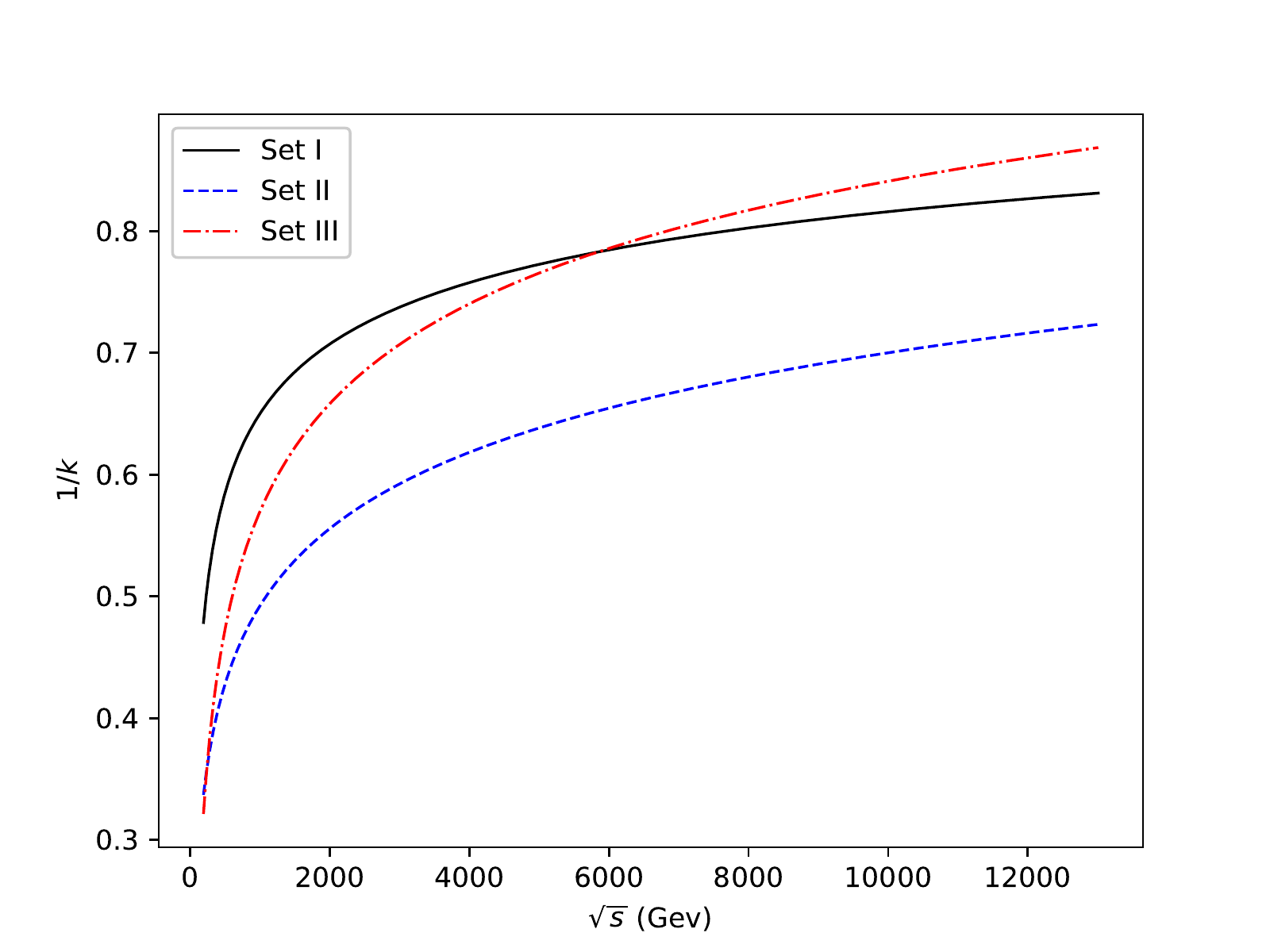}}
}
\caption{$1/k$ calculated with (\ref{1overk})  as a function of the
  collision energy for the three different sets.}
\label{fig3}
\end{figure}

Finally, having fitted $<n>$ and $C_2$ we return to (\ref{1overk}) and
plot $1/k$ as a function of $\sqrt{s}$. The result, shown in Fig.
\ref{fig3} indicates that $1/k$ is an increasing function and there is
no sign of a different behavior. These findings extend the conclusions
found ten years ago in \cite{praza11} to the present energies, which are
two times higher. 

\section{Conclusions}  

We have used the KL (\ref{klmoms}) and BP models (\ref{bpmoms}) to
fit the multiplicity moments measured recently at the LHC.               
The moments in the KL model depend only on $<n>$ and hence  only on 
two free parameter $q_0$ and $\Delta$. Even so it reproduces reasonably    
well the moments of the data of sets I and III (where the model is
indeed expected to be valid as we have mainly gluons in the perturbative
regime). The failure in describing the data of set II is arguably due to
the absence of a component from the fragmentation region. 

In the BP model (\ref{bpmoms}) and assuming a logarithmic growth (\ref{C2fit})
of $C_{2}$ moment, we have been able to reproduce the
multiplicity moments over the wide range of energies for different
rapidity intervals. The input growth of $C_{2}$ with energy can be 
translated through (\ref{1overk}) into a decrease of the parameter
$k$. 
This behavior is consistent with lower energies and does not exhibit the
change predicted in \cite{Gelis09}. 

\vspace{1cm}

\noindent{\bf Acknowledgments:}

We are grateful to the brazilian funding agencies CNPq and CAPES.

\end{document}